\documentclass[aps,prb,showpacs,twocolumn]{revtex4-1}
\usepackage{graphicx,amssymb,amsfonts,amsmath}
\begin{document}

\title{Time-evolution and dynamical phase transitions at a critical time in a system of
one dimensional bosons after a quantum quench
}
\author{Aditi Mitra}
\affiliation{Department of Physics, New York University,
4 Washington Place, New York, New York 10003, USA}
\date{\today}



\begin{abstract}
A renormalization group approach is used to show that a one dimensional system of bosons subject to a
lattice quench exhibits a finite-time dynamical phase transition where
an order parameter within a light-cone increases as a non-analytic function of time after a critical time.
Such a transition is also found for a simultaneous lattice and interaction
quench where the effective scaling dimension of the lattice becomes
time-dependent, crucially affecting the time-evolution of the system.
Explicit results are presented for the time-evolution of the boson interaction parameter
and the order parameter
for the dynamical transition as well as for more general quenches.
\end{abstract}

\maketitle

A fundamental and challenging topic of research is to understand
nonequilibrium strongly correlated systems in general, and how phase transitions occur in such
systems in particular.
While the theory of equilibrium phase transitions
is well developed, and relies heavily on the renormalization group,
the development of an equally powerful approach to study nonequilibrium phase transitions is still in its infancy.
Moreover, in any progress on this topic, it has always appeared that nonequilibrium
phase transitions have one aspect in common
with their equilibrium counterparts, both
occur by adiabatically tuning some parameter of the
system, in the absence or presence of an external
drive, and strictly speaking occur only in the limit of infinite time
(steady-state).~\cite{Mitra06,Mitra08a,Diehl10,Prosen11,Mitra11,Mitra12,Shekhawat11,Torre12}

\begin{figure}
\centering
\includegraphics[totalheight=4cm]{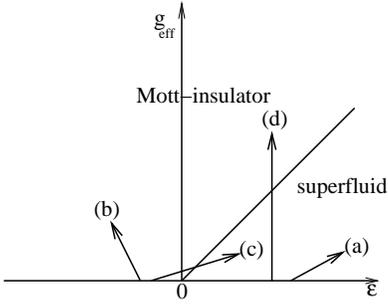}
\caption{The  equilibrium BKT phase diagram. Arrows connect the Hamiltonians before ($H_i$ ) and after ($H_f$) the quench. A dynamical
phase transition is found for case (d).}
\label{pdiag}
\end{figure}

In contrast, here we study
a completely different kind of a nonequilibrium phase transition, one that occurs as a function of time.
Employing a time-dependent renormalization group (RG)
approach we study quench dynamics of interacting one-dimensional
(1D) bosons in a commensurate lattice. This system in equilibrium shows the
Berezinskii-Kosterlitz-Thouless (BKT) transition separating a Mott insulating phase from a superfluid
phase (Fig.~\ref{pdiag}).~\cite{Giamarchibook}
For the nonequilibrium situation we explicitly
show the appearance of a dynamical phase transition where an order-parameter
grows as a non-analytic function of time after a critical time.
Such a behavior has no analog in equilibrium systems.

A dynamical transition in time
was recently identified in the exactly solvable transverse field Ising model where the
Loschmidt echo was found to show non-analytic behavior at a critical time, whereas the
behavior of the order-parameter was analytic.~\cite{Heyl12}
In contrast here we identify a situation
where the order-parameter itself can show non-analyticities as a function of time.
In addition we generalize the
study of dynamical transitions to models
that are not exactly solvable, and to low-dimensions where strong fluctuations negate a
mean-field analysis.~\cite{Sciolla11}

We identify the dynamical transition by studying an order-parameter $\Delta(r,T_m)$
which due to the
quench depends both on position $r$ and a time $T_m$ after the quench. The phase transition
is associated with a non-analytic behavior as a function of time $T_m$ on the value of this order-parameter spatially
averaged within a light cone.
Our results hold relevance not only to experiments in cold-atomic gases where system parameters can be
tuned rapidly in time,~\cite{Bloch08} but also to
conventional solid state materials where time-evolution of an order-parameter may be probed with high precision using
ultra-fast pump-probe~\cite{Fausti11}
and angle resolved photoemission spectroscopy.~\cite{Smallwood12}

We model the 1D Bose gas as a Luttinger liquid,~\cite{Giamarchibook}
\begin{eqnarray}
&&H_i = \frac{u_0}{2\pi}\int dx
\left[K_0\left[\pi \Pi(x)\right]^2 + \frac{1}{K_0}\left[\partial_x \phi(x)\right]^2
\right]
\label{Hidef}
\end{eqnarray}
where $-\partial_x\phi/\pi$ represents the density of the Bose gas, $\Pi$ is the
variable canonically conjugate to $\phi$, $K_0$ is the dimensionless interaction
parameter, and $u_0$ is the velocity of the sound modes.
We assume that the bosons are initially in the ground state of $H_i$.
The system is driven out of equilibrium via
an interaction quench at $t$=$0$ from $K_0 \rightarrow K$, with a commensurate lattice  $V_{sg}$
also switched on suddenly, at the same time as the quench.
This triggers non-trivial time-evolution
from $t >0$ due to a Hamiltonian $H_f=H_{f0} + V_{sg}$, where $H_{f0} = H_i(K_0\rightarrow K)$
and $V_{sg} = -\frac{gu}{\alpha^2} \int dx \cos(\gamma \phi)$, with $g>0$, and
$\Lambda=\frac{u}{\alpha}$ a short-distance cut-off.
We assume that the quench preserves Galilean invariance so that
$u K$=$u_0 K_0$, however this is not critical for either the approach or the result. While
$\gamma$=$2$ for bosons, we keep it general so that the results may be generalized to other 1D systems.

In the absence of the lattice, the system is exactly diagonalizable  in terms of the density modes,
$H_i$=$\sum_{p\neq 0} u_0 |p| \eta_p^{\dagger} \eta_p$ and $H_{f0}$=
$\sum_{p\neq 0}u|p|\beta_p^{\dagger}\beta_p$ where
$\beta,\eta$ are related by a canonical
transformation. This fact has been used to study the dynamics of a Luttinger liquid
exactly, and has revealed
interesting physics arising from a lack of thermalization in the
system.~\cite{Cazalilla06,Iucci09,Perfetto11,Dora11}
To study the system in the presence of the lattice employing RG,
we write the Keldysh action representing the
time-evolution from the initial pure state $|\phi_i\rangle$ (hence an initial density matrix
$\rho =|\phi_i\rangle\langle\phi_i| $) corresponding to the ground state of $H_i$,
$Z_K =Tr\left[\rho(t)\right]= Tr\left[e^{-iH_f t}|\phi_i\rangle\langle\phi_i|e^{i H_f t}\right]$
=$ \int {\cal D}\left[\phi_{cl},\phi_q\right] e^{i \left(S_0 + S_{sg}\right)}$.
$S_0$ is the quadratic part which describes the nonequilibrium Luttinger liquid, which
at a time $t$ after the quench is,~\cite{suppS0}
\begin{eqnarray}
&&S_0 = \int_{-\infty}^{\infty} dx_1 \int_{-\infty}^{\infty} dx_2\int_0^{t} dt_1\int_0^t dt_2
\begin{pmatrix} \phi_{cl}(1) & \phi_q(1)\end{pmatrix}\nonumber \\
&&\begin{pmatrix} 0&&G_A^{-1}(1,2)\\
G_R^{-1}(1,2) && -\left[G_R^{-1} G_K G^{-1}_A\right](1,2)
\end{pmatrix}\begin{pmatrix}
\phi_{cl}(2)\\
\phi_q(2)
\end{pmatrix}\label{S0}
\end{eqnarray}
where $1(2)$=$(x_{1(2)},t_{1(2)})$ and
$\phi_{cl,q}$=$\frac{\phi_-\pm \phi_+}{\sqrt{2}}$
with $-/+$ representing fields that are time/anti-time ordered on the Keldysh contour.~\cite{Kamenevrev}
The lattice potential is given by
$S_{sg}$=$\frac{gu}{\alpha^2}\int_{-\infty}^{\infty}dx_1 \int_0^t dt_1\left[\cos\{\gamma\phi_-(1)\}
-\cos\{\gamma\phi_{+}(1)\}\right]$

We define an order-parameter
$\Delta_m $=$ \langle e^{i m \gamma \phi_{cl}(x,t)}\rangle$ such that in equilibrium $\Delta_{m<1}$
is zero in the gapless phase and non-zero in the gapped phase, and $\Delta_1$ while always non-zero,
is a non-analytic function of $g$.
We will show that after a quench $\Delta_1$ can be a non-analytic function of time.
In order to understand the framework of the RG, let us study
the two-point correlation
function $C_{ab}(1,2)$=$\langle e^{i\gamma \phi_a(1)}e^{-i\gamma \phi_b(2)}\rangle$
($a,b$=$\pm$) for $g$=$0$ but for the
nonequilibrium Luttinger liquid ($K_0 \neq K$).
Denoting $1(2)$=$R+(-)\frac{r}{2},T_m+(-)\frac{\tau}{2}$, $C_{ab}$
depends both on the time-difference $\tau$ as well as the mean time $T_m$ after the quench,
and is translationally invariant in space.~\cite{suppC} $C_{ab}$
depends on three exponents $K_{eq}$=$\frac{\gamma^2 K}{4},
K_{neq}$=$\frac{\gamma^2}{8}K_0\left(1+\frac{K^2}{K_0^2}\right),
K_{tr}$=$\frac{\gamma^2}{8}K_0\left(1-\frac{K^2}{K_0^2}\right)$.
Consider $C_{ab}$ at equal-time ($\tau$=$0$) and unequal positions.
Then for short-times $T_m\Lambda \ll 1$ after the quench but long distances $r \gg u/\Lambda$,
$C_{ab}$ decays in position as a power-law with the exponent $K_{neq}+K_{tr}$=$\gamma^2K_0/4$
(i.e., $C_{ab}\sim r^{-\gamma^2K_0/4}$). Hence
the short time behavior is determined primarily by the initial wave-function.
In contrast, at long-times, $T_m\Lambda \gg 1$, $C_{ab}$ decays as a power-law but with a new exponent
$K_{neq}$ (i.e., $C_{ab}\sim r^{-K_{neq}}$).
$K_{tr}$ governs the transient behavior connecting these limits.
Further, at long times after the quench $T_m\Lambda \gg 1$, $C_{ab}$ also becomes
translationally invariant in time (hence independent of $T_m$).

The RG in equilibrium sums the leading logarithms. We
use the same philosophy to employ RG to study dynamics.
In particular at short times (but long distances), the RG will
resum the logarithms
$\frac{\gamma^2K_0}{4}\ln|r|$ whereas at long times ($T_m\Lambda \gg 1$), it will resum
the logarithms
$K_{neq}\ln\sqrt{(r\pm \tau)^2}$.
Our approach generalizes the use of RG to study quench dynamics near
classical critical points~\cite{Gambassi05,Janssen89} to quantum systems.

{\bf Derivation of RG equations}:
We split the field $\phi_{0,\Lambda}$ into slow ($\phi^<_{0,\Lambda-d\Lambda}$) and fast
($\phi^{>}_{\Lambda-d\Lambda,\Lambda}$) components in momentum space
$\phi_{\pm} = \phi^{<}_{\pm} + \phi^{>}_{\pm}$, and integrate out the fast fields perturbatively in $g$.
Following this we rescale the cut-off back to its original value and rescale position and
time $R,T_m \rightarrow \frac{\Lambda}{\Lambda^{\prime}}(R,T_m)$, where $\Lambda^{\prime} = \Lambda-d\Lambda$.
Following this we write the action as
$S = S_0^{<} +  S_{sg}^{<}+\delta S_0^{<} + \delta S_{T_{eff}}^{<} + \delta S_{\eta}^{<}$
where $S_{0}^{<}$ is simply the quadratic action corresponding to $H_{0f}$
with the rescaled variables, $S^<_{sg}$ is the rescaled action due to the lattice, while
$\delta S^<_{0,T_{eff},\eta}$ are corrections to ${\cal O}(g^2)$.
\begin{eqnarray}
&&S^<_{sg} = g\left(\frac{\Lambda}{\Lambda^{\prime}}\right)^2\int_{-\infty}^{\infty}dR
\int_0^{t\Lambda} dT_m
\left[\cos\gamma\phi_-^<(R,T_m) \right. \nonumber \\
&&\left. - \cos\gamma\phi_+^<(R,T_m)\right]
e^{-\frac{\gamma^2}{4}\langle\left[\phi_{cl}^>(T_m)\right]^2\rangle}\label{dSg}\\
&&\delta S^<_0 = \frac{g^2\gamma^2}{2}\frac{d\Lambda}{\Lambda}
\int_{-\infty}^{\infty} dR\int_0^{t\Lambda/\sqrt{2}} dT_m
\left[ -I_R(T_m)
\left(\partial_R \phi_{cl}^<\right)\right. \nonumber \\
&&\left. \times \left(\partial_R \phi_{q}^<\right)
 - I_{ti}(T_m)
\left(\partial_{T_m} \phi_{cl}^<\right) \left(\partial_{T_m} \phi_{q}^<\right)\right]\label{dS0}\\
&& \delta S_{T_{eff}}^{<}= \frac{ig^2\gamma^2}{2}\frac{d\Lambda}{\Lambda}
\int_{-\infty}^{\infty}\!\!\!\! dR\!\!\int_0^{t\Lambda/\sqrt{2}}\!\!\! \!\!\!\!\! dT_m
\left(\phi_q^{<}\right)^2I_{T_{eff}}(T_m)\\
&& \delta S_{\eta}^{<} = -\frac{g^2\gamma^2}{2}\frac{d\Lambda}{\Lambda}
\int_{-\infty}^{\infty}\!\!\!\! dR\!\!\int_0^{t\Lambda/\sqrt{2}}\!\!\! \!\!\!\!\!dT_m
\phi_q^<\left[\partial_{T_m}\phi_{cl}^<\right] I_{\eta}(T_m)
\end{eqnarray}
Eq.~(\ref{dSg}) shows that the scaling dimension of the lattice depends on $\langle\left[\phi_{cl}^>(T_m)\right]^2\rangle$,
and in particular is time-dependent. To leading order ($g$=$0$)
$\frac{\gamma^2}{4}\langle\left[\phi_{cl}^>(T_m)\right]^2\rangle = \frac{d\Lambda}{\Lambda}\left[K_{neq}+
\frac{K_{tr}}{1+ \left(2T_m \Lambda\right)^2}\right]$.
$\delta S^{<}_{0}$ shows that the quadratic part of the action acquires corrections
which are also time-dependent.~\cite{supp1}
$\delta S^{<}_{\eta,T_{eff}}$ indicates the generation of new terms such as a time-dependent
dissipation ($\eta$) and a noise $\left(\eta T_{eff}\right)$ whose physical meaning is the generation of
inelastic scattering processes which will eventually relax the distribution
function.~\cite{Mitra11,Mitra12} These time dependent corrections lead to RG equations
which depend on time $T_m$ after the quench.
Defining $\frac{d\Lambda}{\Lambda}$=$ \frac{\Lambda-\Lambda^{\prime}}{\Lambda}$=$d\ln(l)$,
$I_{u,K}$=$I_R \pm I_{ti}$~\cite{supp1}, and
dimensionless variables
$T_m \rightarrow T_m \Lambda, \eta \rightarrow \eta/\Lambda, T_{eff}\rightarrow T_{eff}/\Lambda$
the RG equations are,
\begin{eqnarray}
\frac{dg}{d\ln{l}} = g\left[2-\left(K_{neq} + \frac{K_{tr}}{1+4 T_m^2}\right)\right]
\label{rg1}\\
\frac{dK^{-1}}{d\ln{l}} = \frac{\pi g^2 \gamma^2}{8}I_K(T_m)\label{rg2}\\
\frac{dT_m}{d\ln{l}} = -T_m \label{rg3}\\
\frac{1}{Ku}\frac{du}{d\ln{l}} = \frac{\pi g^2 \gamma^2}{8}I_u(T_m)\label{rg4}\\
\frac{d\eta}{d\ln{l}} = \eta + \frac{\pi g^2\gamma^2 K}{4}I_{\eta}(T_m)\label{rg5}\\
\frac{d(\eta T_{eff})}{d\ln{l}} = 2 \eta T_{eff} + \frac{\pi g^2\gamma^2 K}{8}I_{T_{eff}}(T_m)
\label{rg6}
\end{eqnarray}
Note that $T_m$ not only acts as an inverse cut-off
in that modes of momenta $\Lambda <1/T_m$ dominate the physics at a time $T_m$,~\cite{Mathey09,Vosk12} 
it also governs the crossover from a short time behavior where the physics is determined primarily
by the initial state, and a long time behavior characterized by a new
nonequilibrium fixed point. This crossover is most easily seen from Eq.~(\ref{rg1})
where the scaling dimension of the lattice $\epsilon(T_m) = \left[K_{neq} + \frac{K_{tr}}{1+4 T_m^2}-2\right]$
depends on time as follows:
at short times ($T_m\ll 1$) it is $(-2+\frac{\gamma^2K_0}{4})$, and hence depends on
the initial wave-function, at long times ($T_m\gg 1$), a nonequilibrium scaling dimension
$(-2+K_{neq})$ emerges.

Above, $I_{K,u,\eta,T_{eff}}$ reach steady state
values at $T_m\gg 1$, whereas for short times,
they vanish as $T_m\rightarrow 0$ as expected since the effect of the lattice potential vanishes at
$T_m$=$0$. For example, at short times $I_{K}\sim {\cal O}(T_m^2)$.~\cite{suppIK,noteLE}
Eqns.~(\ref{rg2}) and~(\ref{rg4}) represent renormalization of the interaction
parameter and the velocity. The effects of the  latter being small will be neglected, and in what follows
we set $u$=$1$.
Eqns~(\ref{rg5}),~(\ref{rg6}) show the generation of dissipation and noise that represent inelastic scattering between
bosonic modes.~\cite{Mitra11,Mitra12}

In what follows we
do an analysis for a time $T_m < 1/\eta$ where $1/\eta$ is the time in which the distribution function
first begins to change due to inelastic scattering.
A perturbative calculation~\cite{Mitra11,Mitra12}
shows that for small quenches ($|K_0-K| \rightarrow 0$), and at steady-state, $\eta \sim g^2 (K_0-K)^4$.
Since $\eta \ll 1$,
one may easily be in the regime of $T_m \gg 1$ but $T_m\eta \ll 1$ so that inelastic
scattering may be neglected. At these times, and in what follows we will only use
equations~(\ref{rg1}),~(\ref{rg2}) and~(\ref{rg3}).

The behavior of the system is very different depending upon $K_0,K,g$.
We discuss four cases (see Fig~\ref{pdiag}). Case (a) is when the periodic potential is irrelevant at all
times after the quench,
case (b) is when the periodic potential is always relevant,
case (c) is when the periodic potential is relevant at short times, and irrelevant at long times,
while case (d) is when the periodic potential is irrelevant at short times and relevant at long times. For case (d)
we show that an order-parameter behaves in a discontinuous way in time. There is a critical
time $T_m^*$ after which the order-parameter begins to increase as a non-analytic function of time indicating a
dynamical phase transition. In contrast,
for case (c), the  behavior of the order-parameter is analytic in time.

We use $\epsilon_0,g_0,T_{m0}, \Lambda_0$ to denote bare physical values. 
From Eq.~(\ref{rg2}) we define an effective-interaction
$g_{eff}(T_m)$=$g\sqrt{\frac{\pi\gamma^2}{8}}\sqrt{I_K(T_m)}\frac{\gamma K}{2}
\sqrt{\frac{K}{K_0}}$ where $g_{eff}$ goes to zero as $T_{m}\rightarrow 0$ and reaches a steady state value for
$T_{m}\gg 1$.
Physically this implies that at short times the particles have not had sufficient time to interact,
therefore however large $g$ may be, any renormalization effects due to interactions is vanishingly small.
The time-dependence of $g_{eff}(T_m)$ and $\epsilon(T_m)$ will be important for the results.

{\bf Case (a), periodic potential always irrelevant~\cite{suppRG}}: This occurs for $\epsilon(T_m) >0$
and $g_{eff}$ not too large (a condition to be made more precise when discussing case (d)).
Here the periodic potential renormalizes to zero, and one recovers a gapless theory which eventually looks
thermal at $T_m \gg  1/\eta$.~\cite{Mitra11, Mitra12}
The RG predicts how quantities renormalize in time and in particular
shows that at long times the steady-state state is approached as a power-law with a non-universal exponent
$\epsilon^* \xrightarrow{\Lambda_0T_{m0}\gg 1} A + {\cal O}\left(\frac{1}{\Lambda_0 T_{m0}}\right)^{2 A}$, where
$A = \sqrt{\epsilon_0^2(\infty)-g_{eff,0}^2(\infty)}$.

{\bf Case (b), periodic potential always relevant~\cite{suppRG}}: This occurs for $\epsilon(T_m)<0$. Thus
we are always in the strong coupling regime. Here we integrate the RG equations upto a scale
$l^*(T_m)$ where the renormalized coupling is {\cal O}(1). Beyond this
scale our RG equations are not valid, however the advantage of the bosonic theory is that at strong-coupling
$g\cos(\gamma \phi)\simeq g(1- \gamma^2\phi^2/2 + \ldots)$ so that $\sqrt{g}$ may be identified with a gap.
The physical gap/order-parameter is then given by $\Delta = \sqrt{g}/l^* = 1/l^*$.
Since $l^*(T_m)$ depends on time, it tells us how the order-parameter evolves in time.~\cite{suppRG2}

Let us first consider short times $T_{m0}\Lambda_0\ll 1$. Here perturbation theory is valid, and gives
$\Delta_1 \sim g_0 T_{m0}^2$, a result which is consistent with a lattice quench at the
the exactly solvable Luther-Emery point~\cite{Iucci10}.
At long times after the quench, the scaling dimension is $2-K_{neq}$. Provided that
$\Delta T_{m0}\gg 1$, we find the steady-state order-parameter, $\Delta_{ss}= \left(g_{eff,0}\right)^{\frac{1}{2-K_{neq}}}$.
Compare this with the order-parameter in the ground state of $H_f$~\cite{Giamarchibook} $\Delta_{eq}$=$\left(g_{eff,0}\right)^{\frac{1}{2-K_{eq}}}$.
Since $K_{neq}>K_{eq}$ and $g_{eff,0}\ll 1$, the order-parameter at long times after the quench is always
smaller than the order-parameter in equilibrium.
The RG equations may also be solved at intermediate times~\cite{suppRG} $\frac{1}{\Lambda_0}\ll T_{m0}\ll \frac{1}{\Delta}$. Here we find,
$\Delta =
\left[
\left(\Lambda_0T_{m0}\right)^{\frac{\gamma^2K_0}{4}-K_{neq}}g_{eff,0}\right]^{\frac{1}{2-\frac{\gamma^2K_0}{4}}}
$.
Thus at intermediate times the gap decreases with time if $K_{neq}> \frac{\gamma^2 K_0}{4}$,
or increases with time for the reverse case. For $K_0$=$K$
this intermediate time power-law dynamics is absent.
\begin{figure}
\centering
\includegraphics[totalheight=5cm]{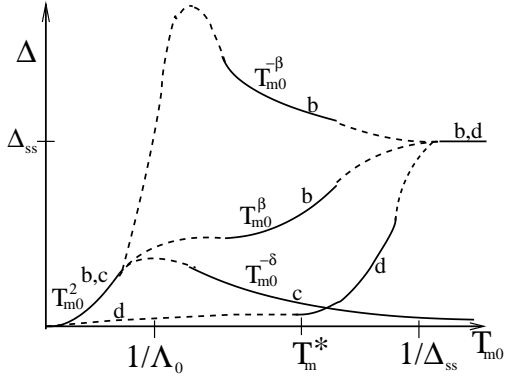}
\caption{Time evolution  of the order-parameter after the quench for cases (b), (c) and (d).
Solid lines show a short time behavior ($T_{m0}\ll \frac{1}{\Lambda_0}$), an intermediate
time asymptotics ($ \frac{1}{\Lambda_0} \ll T_{m0} \ll \frac{1}{\Delta_{ss}}$) and a
long time behavior $T_{m0}\gg \frac{1}{\Delta_{ss}}$. At intermediate times the order-parameter
increases as $T_{m0}^{\beta}$ (decreases as $T_{m0}^{-\beta}$)
when $K_{neq}<\frac{\gamma^2 K_0}{4}$
($K_{neq}>\frac{\gamma^2 K_0}{4}$) for case (b) and eventually reaches a steady-state value
$\Delta_{ss}$.
For case (c) the order-parameter decreases for $T_{m0} \gg \frac{1}{\Lambda_0}$
as $\Delta \propto T_{m0}^{-\delta}$.
For case (d) the order-parameter increases after time $T_{m}^*$ in a non-analytic manner in time (Eq.~(\ref{gapdyn})).
$\beta$=$\theta\left(|\frac{\gamma^2K_0}{4}-K_{neq}|\right)$, $\delta=1+A\theta$, $\theta$=$\frac{1}{2-\frac{\gamma^2K_0}{4}}$,
and $A = \sqrt{\epsilon_0^2-g_{eff,0}^2}$.
Dashed lines are a guide to the eye for the crossover regimes.}
\label{gaptimev}
\end{figure}

{\bf Case (c), periodic potential relevant at short times, and irrelevant at long times~\cite{suppRG}}.
This occurs when $\epsilon(T_m)$ changes sign from negative to positive and $g_{eff}$ is not too large.
Here the short time behavior is the same as Case (b), however at long times,
the order-parameter decreases with time as $\Delta \sim
 \left(\frac{1}{\Lambda_0 T_{m0}}\right)^{1+\frac{A}{2-\frac{\gamma^2K_0}{4}}}$.
Fig.~\ref{gaptimev} summarizes the behavior of the order-parameter for cases (b), (c) and (d), the last case to be discussed next. The non-monotonic
dependence of the order-parameter in time is due to the time-dependence of the scaling dimension which physically
leads to a situation where quantum fluctuations are enhanced (suppressed) at a later time for
$\epsilon(T_m=\infty)>\epsilon(T_m=0)$
($\epsilon(T_m=\infty)<\epsilon(T_m=0)$), causing the order-parameter to decrease (increase).

{\bf Case (d), periodic potential irrelevant at short times and relevant at long times~\cite{suppRG}}:
This occurs under two conditions. Either $\epsilon(T_m)$
changes sign from positive to negative during the time-evolution, or
$\epsilon(T_m)$ is always positive,  but
$g_{eff}(T_m)$ becomes sufficiently large at some time $T_m^*$. The latter includes the case of a pure lattice
quench ($K_0$=$K$). For either condition, the RG treatment, which neglects the effect of irrelevant
operators shows that at long times, the order-parameter reaches a steady state value,
while at short times it is zero.
This indicates a non-analytic behavior at a critical time $T_{m}^*$.

Fig.~\ref{pdiag} contrasts
case (d) with the previous cases considered where the order-parameter behaved analytically.
The renormalized interaction parameter $g_{eff}(T_m)$
is vanishingly small right after the quench. For case (b), since infinitesimally small $g_{eff}$ is a relevant perturbation,
an order-parameter starts growing immediately after the quench. On the other hand for a quench corresponding to
case (d), Fig.~\ref{pdiag} shows that $g_{eff}$ has to be larger than a critical value in order to be in the Mott-phase.
Thus one has to wait some finite time before which renormalization effects become large enough for an order-parameter to grow.
We now discuss this physics in a more quantitative manner, and for simplicity, consider only the case
of the pure lattice quench.

Let us suppose $T_{m0}\Lambda_0 \gg 1$. Here the RG equations are solved in two steps, one for
$1<l<T_{m0}\Lambda_0$, and the other for $T_{m0}\Lambda_0<l$.
For  the first step, since $I_K$ varies slowly at long times, eventually reaching a steady-state value,
we may assume $\frac{1}{2}|\frac{dI_K}{d\ln{l}}|\ll |\frac{d\ln{g}}{d\ln{l}}|$. Thus the RG equations
are the conventional ones of the equilibrium BKT transition
$\frac{dg_{eff}}{d\ln{l}}$=$-g_{eff}\epsilon, \frac{d\epsilon}{d\ln{l}}$=$-g^2_{eff}$.
For the second step,
($\Lambda_0 T_{m0}<l$), since $I_K \sim T_m^2$, the RG equations become $\frac{dg_{eff}}{d\ln{\bar{l}}} = -g_{eff}\epsilon, \frac{d\epsilon}{d\ln{\bar{l}}}=-\frac{g^2_{eff}}{\bar{l}^2}$, where $\bar{l}=\frac{l}{\Lambda_0 T_{m0}}$.
The solution shows that there is a critical time $T_m^*$ such that $g_{eff}$ is irrelevant before this time, and is a relevant perturbation after this time. We find,
\begin{eqnarray}
\!\!\Lambda_0 T_m^* \!\!= e^{\frac{1}{D}\arctan\left(\frac{\epsilon_0}{D}\right)-\frac{1}{D}\arctan{\frac{D}{2}}},
D^2 = g^2_{eff,0}-\epsilon_0^2
\end{eqnarray}
The deeper one quenches into the Mott-phase, the shorter is $T_m^*$. Moreover, $T_m^*$ is longest along the critical
line $g_{eff,0}$=$\epsilon_0$.
By identifying a length-scale at which $g_{eff}(\bar{l}^*) \sim 1$, we find that the order-parameter grows as
\begin{eqnarray}
&&\!\!\Delta \!\!\sim \Delta_{smooth} +\nonumber \\
&&\theta(T_{m0}-T_m^*)\frac{1}{\Lambda_0 T_{m0}}\!\!\left[g_{eff}(l=\Lambda_0T_{m0})\right]^{\frac{f_2}{(T_{m0}-T_m^*)}}
\label{gapdyn}
\end{eqnarray}
where $f_2=\frac{1}{|\frac{d\epsilon(l=T_m)}{dT_m}|_{T_m=T_m^*}|}$ and $\Delta_{smooth}$ is a background contribution arising
from irrelevant operators whose effects may  be treated perturbatively. Thus while $\Delta$ is always non-zero after the quench
due to the presence of irrelevant terms,
due to the relevant terms, it increases as a non-analytic function of time after a critical time.

An important question concerns the spatial variation of the order-parameter. Quenches in gapless systems
are associated with light-cone dynamics where two points a position $R$ apart get correlated after a time
$T_m\sim R$.~\cite{Calabrese06} For our case any two points separated  by $R > T_m$ will behave primarily like
the initial state with power-law correlations in position determined by $K_0$. The predictions for
the order-parameter made above is for a region within a light cone $R < T_m$. The dynamical
transition at $T_m^*$ is associated with the appearance of order in regions of size $R^*\sim T_m^*$,
after which the ordered regions will begin to grow in size.

In summary, employing RG we have identified a novel dynamical phase transition in a strongly correlated
system where an order-parameter grows as a non-analytic function of time
after a critical time (Eq.~(\ref{gapdyn})). The order parameter shows rich dynamics
both at the transition as well as for more general quenches (Fig.~\ref{gaptimev}).
Identifying similar dynamical transitions in
higher dimensions where thermal fluctuations are less effective in destroying order is an important direction of research.

{\sl Acknowledgements:} The author gratefully acknowledges helpful discussions with
I. Aleiner, B. Altshuler, E. Dalla Torre, E. Demler, P. Hohenberg, A. Millis, E. Orignac and M. Tavora.
This work was supported by NSF-DMR (1004589) and NSF PHY05-51164.


\begin{center}
\underline{\Large{Supplementary Material}}
\end{center}
\section{Elements of the quadratic action and outline of the RG procedure}
The action for the non-equilibrium Luttinger liquid is
\begin{eqnarray}
&&S_0 = \int_{-\infty}^{\infty} dx_1 \int_{-\infty}^{\infty} dx_2\int_0^{t} dt_1\int_0^t dt_2
\begin{pmatrix} \phi_{cl}^*(1) & \phi_q^*(2)\end{pmatrix}\nonumber \\
&&\begin{pmatrix} 0&&G_A^{-1}(1,2)\\
G_R^{-1}(1,2) && -\left[G_R^{-1} G_K G^{-1}_A\right](1,2)
\end{pmatrix}\begin{pmatrix}
\phi_{cl}(2)\\
\phi_q(2)
\end{pmatrix}\label{S0s}
\end{eqnarray}
Denoting $1=x_1,t_1; 2 = x_2,t_2$
\begin{eqnarray}
&&G_{R,A}^{-1}(1,2) \nonumber \\
&&= -\delta (x_1-x_2) \delta(t_1-t_2)\frac{1}{\pi K u}
\left[\partial_{t_1\pm i\delta}^2 - \partial_{x_1}^2\right]
\end{eqnarray}
while
\begin{eqnarray}
&&G^K(xt_1,yt_2)
= -i\frac{K_0}{2}\left(1+\frac{K^2}{K_0^2}\right)\int_0^{\infty}\frac{dp}{p}e^{-\alpha p}\nonumber \\
&&\cos(up(x-y))
\cos(u|p|(t_1-t_2)) \nonumber \\
&&-i\frac{K_0}{2}\left(1-\frac{K^2}{K_0^2}\right)\int_0^{\infty}\frac{dp}{p}e^{-\alpha p}\nonumber \\
&&\cos(up(x-y))
\cos(u|p|(t_1+t_2))
\end{eqnarray}
where
\begin{eqnarray}
\Lambda = \frac{u}{\alpha}
\end{eqnarray}
is a short-distance cut-off.

In doing the RG, the fields $\phi$ are split into slow fields ($\phi^<$) that have
Fourier modes in momentum space between $0,\Lambda-d\Lambda$, and fast fields ($\phi^>$)
that have Fourier modes in momentum space between $\Lambda-d\Lambda,\Lambda$.
Thus $\phi= \phi^{<} + \phi^>$.
The fast modes are integrated out perturbatively in the periodic potential. In doing so
we use the fact that the correlator for the field $\phi$ ($G_{0,\Lambda}$) is related to
the correlators for the slow ($G^<_{0,\Lambda-d\Lambda}$) and fast fields
($G^>_{\Lambda-d\Lambda,\Lambda}$) as follows $G_{0,\Lambda} = G^<_{0,\Lambda-d\Lambda}
+ G^>_{\Lambda-d\Lambda,\Lambda}$, so that~\cite{Nozieres87}
\begin{eqnarray}
G^>_{\Lambda-d\Lambda,\Lambda}=d\Lambda \frac{dG_{0,\Lambda}}{d\Lambda}
\end{eqnarray}
Following this we rescale the cut-off back to its original value and in the process rescale position and
time to $x_i,t_i \rightarrow \frac{\Lambda}{\Lambda^{\prime}}(x_i,t_i)$ where $\Lambda^{\prime}=\Lambda-d\Lambda$.

\section{Expression for $C_{+-}(r,T_m,\tau)$}

The two-point correlator defined as
\begin{eqnarray}
C_{+-}(r,T_m,\tau) = \langle e^{i\phi_+(r,T_m+\tau/2)}e^{-i\phi_-(0,T_m-\tau/2)}\rangle
\end{eqnarray}
is for $g=0$ given by
\begin{eqnarray}
&&C_{+-}(r,T_m,\tau)
=\left[\frac{\alpha}{\sqrt{\alpha^2 + (u\tau+r)^2}}
\frac{\alpha}{\sqrt{\alpha^2 + (u\tau-r)^2}}\right]^{K_{neq}}
\nonumber \\
&& \times
\left[\frac{\sqrt{\alpha^2 + \{2u(T_m+\tau/2)\}^2}}{\sqrt{\alpha^2 + (2uT_m+r)^2}}\frac{\sqrt{\alpha^2
+ \{2u (T_m-\tau/2)\}^2}}{\sqrt{\alpha^2 + (2uT_m-r)^2}}\right]^{K_{tr}}\nonumber \\
&&\times e^{-i K_{eq}\left[
\tan^{-1}\left(\frac{u\tau+r}{\alpha}\right)
+ \tan^{-1}\left(\frac{u\tau-r}{\alpha}\right)
\right]}\label{GKtr}
\end{eqnarray}

\section{Expressions for $I_R,I_{ti},I_{\eta},I_{T_{eff}},I_{K},I_u$}
\begin{eqnarray}
I_{R}(T_m\Lambda)&&= -\int_{-\infty}^{\infty} d\bar{r} \int_{0}^{2T_m\Lambda } d\bar{\tau} \bar{r}^2\nonumber \\
&&{\rm Im}\left[B(\bar{r},T_m\Lambda,\bar{\tau})\right]\\
I_{ti}(T_m\Lambda)&&= -\int_{-\infty}^{\infty} d\bar{r} \int_{0}^{2T_m\Lambda} d\bar{\tau} \bar{\tau}^2\nonumber \\
&&{\rm Im}\left[B(\bar{r},T_m\Lambda ,\bar{\tau})\right]\\
I_{T_{eff}}(T_m)&&= \int_{-\infty}^{\infty} d\bar{r} \int_{-2T_m\Lambda}^{2T_m\Lambda} d\bar{\tau}\nonumber \\
&&{\rm Re}\left[
B(\bar{r},T_m\Lambda,\bar{\tau})\right]\\
I_{\eta}(T_m\Lambda)&&= -\int_{-\infty}^{\infty} d\bar{r} \int_{-2T_m\Lambda}^{2T_m\Lambda} d\bar{\tau}  \bar{\tau}\nonumber \\
&&{\rm Im}\left[
B(\bar{r},T_m\Lambda,\bar{\tau})\right]\\
I_{u}(T_m\Lambda)&&= -\int_{-\infty}^{\infty} d\bar{r} \int_{0}^{2T_m\Lambda } d\bar{\tau} \left(\bar{r}^2+\bar{\tau}^2\right)
\nonumber \\
&&{\rm Im}\left[B(\bar{r},T_m\Lambda,\bar{\tau})\right]\\
I_{K}(T_m\Lambda)&&= -\int_{-\infty}^{\infty} d\bar{r} \int_{0}^{2T_m\Lambda} d\bar{\tau}\left(\bar{r}^2- \bar{\tau}^2\right)\nonumber\\
&&{\rm Im}\left[B(\bar{r},T_m\Lambda ,\bar{\tau})\right]\label{IK}
\end{eqnarray}
where ${\rm Re}[B]=(B+B^*)/2$, ${\rm Im}[B]$=$(B-B^*)/(2i)$, and
\begin{eqnarray}
B(\bar{r},T_m\Lambda,\bar{\tau})=C_{+-}(\bar{r},T_m\Lambda,\bar{\tau})F(\bar{r},T_m\Lambda,\bar{\tau})
\end{eqnarray}
with $C_{+-}(\bar{r},T_m\Lambda,\bar{\tau})=\langle e^{i\phi_+(\bar{r},\bar{\tau}+T_m\Lambda/2)}e^{-i\phi_-(0,\bar{\tau}-T_m\Lambda/2)} \rangle$. This quantity within leading order in perturbation theory is given by Eq.~\ref{GKtr} which we rewrite in dimensionless
variables,
\begin{eqnarray}
&&C_{+-}(\bar{r},T_m\Lambda,\bar{\tau})
=\left[\frac{1}{\sqrt{1 + (\bar{\tau}+\bar{r})^2}}
\frac{1}{\sqrt{1 + (\bar{\tau}-\bar{r})^2}}\right]^{K_{neq}}
\nonumber \\
&&
\left[\frac{\sqrt{1 + \{2(T_m\Lambda+\bar{\tau}/2)\}^2}}{\sqrt{1 + (2T_m\Lambda+\bar{r})^2}}\frac{\sqrt{1
+ \{2 (T_m\Lambda-\bar{\tau}/2)\}^2}}{\sqrt{1 + (2T_m\Lambda-\bar{r})^2}}\right]^{K_{tr}}\nonumber \\
&&\times e^{-i K_{eq}\left[
\tan^{-1}\left(\bar{\tau}+{\bar r}\right)
+ \tan^{-1}\left(\bar{\tau}-\bar{r}\right)
\right]}
\end{eqnarray}
while $F$ is given by
\begin{eqnarray}
&&F(\bar{r},T_m\Lambda,\bar{\tau}) = K_{neq}\left[\frac{1}{1 + \left(\bar{\tau}+\bar{r}\right)^2} +
\frac{1}{1 + \left(\bar{\tau}-\bar{r}\right)^2}\right]\nonumber \\
&&+K_{tr}\left[\frac{1}{1 + \left(2T_m\Lambda +\bar{r}\right)^2} + \frac{1}{1 +
\left(2T_m\Lambda-\bar{r}\right)^2}\right]\nonumber \\
&&- i K_{eq}\left[\frac{\bar{\tau}+\bar{r}}{1 + \left(\bar{\tau}+\bar{r}\right)^2}
+ \frac{\bar{\tau}-\bar{r}}{1 + \left(\bar{\tau}-\bar{r}\right)^2}\right]
\end{eqnarray}
and
\begin{eqnarray}
&&K_{eq}=\frac{\gamma^2 K}{4}\\
&&K_{neq}=\frac{\gamma^2}{8}K_0\left(1+\frac{K^2}{K_0^2}\right)\\
&&K_{tr}=\frac{\gamma^2}{8}K_0\left(1-\frac{K^2}{K_0^2}\right)
\end{eqnarray}
We will often use dimensionless variables $T_m \leftrightarrow T_m\Lambda$.

\section{Short time behavior of $I_K$}
$I_K$ depends on the two-point function $C_{+-}$ which may be computed explicitly within
perturbation theory. We discuss its short-time behavior ($T_{m}\Lambda \ll 1$) here. In this case
the integrand in Eq.~\ref{IK} may be Taylor expanded in ${\tau}$ so that
\begin{eqnarray}
&&I_K(T_m\ll 1) \simeq \nonumber\\
&&2 K_{eq} \int_{-\infty}^{\infty}dr \int_0^{2T_m} d\tau\Biggl\{ r^2 \left(\frac{1}{1+r^2}\right)^{\frac{\gamma^2K_0}{4}} \times \Biggr.\nonumber \\
&&\Biggl.
\times \left[\frac{1-r^2}{(1+r^2)^2} + \frac{\gamma^2 K_0}{4}\frac{2}{(1+r^2)^2}\right]\tau + {\cal O}(\tau^3)\Biggr\}\nonumber \\
&&= 4 \sqrt{\pi} K_{eq}\left(\frac{\gamma^2K_0}{4}-1\right)\frac{\Gamma(\frac{\gamma^2K_0}{4}-\frac{1}{2})}
{\Gamma(1+\frac{\gamma^2 K_0}{4})}T_m^2 + {\cal O}(T_m^4)\nonumber \\ \label{IKshort}
\end{eqnarray}
Note that $I_K$ generically behaves as $I_K \sim T_m^2$ at short times.
However at the Luther-Emery point $\frac{\gamma^2K_0}{4}=1$ so that the term of ${\cal O}(T_m^2)$ vanishes, and the
leading behavior is $I_K \sim {\cal O}(T_m^4)$.

It is important to note that the Luther-Emery point is quite deep in the
Mott phase, and is accessible perturbatively  in $g$ only at short times. At
long times, the slow power-law decay in time (and in space) leads to infrared singularities that
makes perturbation theory break-down. In this paper we do not
attempt to discuss the long-time behavior at the Luther-Emery point.

\section{Solution of the RG equations}
If we are interested in times that are such that $T_m < 1/\eta$, where
$1/\eta$ is the time in which the distribution function of the bosons changes considerably due to inelastic scattering,
we may simplify the RG equations to
\begin{eqnarray}
\frac{dg}{d\ln{l}} = g\left[2-\left(K_{neq} + \frac{K_{tr}}{1+4 T_m^2}\right)\right]
\label{rg1s}\\
\frac{dK^{-1}}{d\ln{l}} = \frac{\pi g^2 \gamma^2}{8}I_K(T_m)\label{rg2s}\\
\frac{dT_m}{d\ln{l}} = -T_m \label{rg3s}
\end{eqnarray}
We will use the notation $g_0,T_{m0},\Lambda_0$ as the bare physical values.
Let us define the scaling dimension of the lattice as
\begin{eqnarray}
\epsilon(T_m) = \left[-2+\left(K_{neq} + \frac{K_{tr}}{1+4 T_m^2}\right)\right]
\end{eqnarray}
We now consider four cases separately:
Case (a) is when $\epsilon(T_m) >0$ at all times. Case (b) is when $\epsilon(T_m) <0$ at all times.
Case (c) is when $\epsilon(T_m) <0$ at short times and $\epsilon(T_m)>0$ at long times. Case (d) corresponds to the
dynamical transition and occurs for
two possible cases, one where
$\epsilon(T_m)>0$ at short times and $\epsilon(T_m)<0$ at long times. The second is
when $\epsilon(T_m) >0$ at all times, but $g_{eff}(T_m)$ becomes sufficiently large at long times.

While one may always solve the
equations ~\ref{rg1s},~\ref{rg2s},~\ref{rg3s} numerically, we can also obtain analytic solutions in two limits, one for short times $T_{m0}\Lambda_0 \ll 1$ and the other for long times $T_{m0}\Lambda_0 \gg 1$. In these two limits a simplification
comes from the fact that $I_K(T_m)$ for $T_m \gg 1$ changes very slowly,
as a power-law with $T_m$, whereas at short times it changes as $T_{m}^2$. For long times,
$g$ changes more rapidly with $l$ than $I_K$ does, so we will neglect the $l$ dependence of $I_K$ at long times. However
at short times, we will retain its $l$ dependence via $I_K(T_m\ll 1) = c T_{m0}^2\Lambda^2/l^2$.

\subsection{Case (a)}
Case (a) is when $\epsilon(T_m) >0$ at all times, and $g$ is not too large (this statement will be made
more precise when discussing Case(d)). In this case,
the periodic potential is always irrelevant.

Let us first consider the short time solution. Here
$\epsilon = \frac{\gamma^2K_0}{4} - 2$. Further we write $I_K(T_m\ll 1) = c T_m^2$. Thus the RG equations become
\begin{eqnarray}
\frac{dg}{d\ln{l}}=g\left(2-\frac{\gamma^2 K_0}{4}\right)\\
\frac{dK^{-1}}{d\ln{l}}\simeq \frac{\pi g^2 \gamma^2}{8}\left( \frac{cT_{m0}^2\Lambda_0^2}{l^2}\right)
\end{eqnarray}
Solving this upto $l \rightarrow \infty$ (which is justified as
$g$ decays rapidly with $l$), the solution is $g^*=0$ and
$\frac{1}{K^*} = \frac{1}{K} + \frac{\pi^2\gamma^2g_0^2 c T_{m0}^2 \Lambda_0^2}{16\left(\frac{\gamma^2K_0}{4}-1\right)}
\simeq \frac{1}{K} + {\cal O}(T_{m0}^2\Lambda_0^2)$. Thus the renormalized interaction parameter $K^*$
changes quadratically with time at short times from its bare value of $K$.

For long times $T_{m0}\Lambda_0\gg 1$,
the RG equations have to be integrated in two
steps, one for $T_m \gg 1$ where the RG equations become $\frac{dg}{d\ln{l}}$=$ g \left(2-K_{neq}\right),
\frac{d K}{d\ln{l}}$=$-\frac{\pi g^2\gamma^2 K^2}{8}I_K(T_{m0}\Lambda_0/l)$. Define $\epsilon = K_{neq}-2$
where $d\epsilon = \frac{\gamma^2}{4}\frac{K}{K_0}dK$.
We will now use the fact
that the $l$ dependence of $I_K(T_{m0}\Lambda_0/l)$ is much weaker than that of $g$ at long times i.e,
($\frac{1}{2}|\frac{d \ln{I_K}}{d\ln{l}}| \ll |\frac{d\ln{g}}{d\ln{l}}|$).
We also define an effective-interaction,
\begin{eqnarray}
g_{eff}(T_{m0}\Lambda_0)=g\sqrt{\frac{\pi\gamma^2}{8}}\sqrt{I_K(T_{m0}\Lambda_0)}\frac{\gamma K}{2}
\sqrt{\frac{K}{K_0}}
\end{eqnarray}

In terms of these new variables, in the long time limit, the RG equations are
\begin{eqnarray}
\frac{d g_{eff}}{d\ln{l}}=-\epsilon g_{eff}\\
\frac{d\epsilon}{d\ln{l}}=- g_{eff}^2
\end{eqnarray}
The solution of the above equations are well known~\cite{Giamarchibook},
\begin{eqnarray}
\epsilon(\ln{l}) = \frac{A}{\tanh\left[A\ln{l} + \tanh^{-1}\left(\frac{A}{\epsilon_0}\right)\right]}\label{esol1}\\
g\left(\ln{l}\right) = \frac{A}{\sinh\left[A\ln{l} + \tanh^{-1}\left(\frac{A}{\epsilon_0}\right)\right]}\label{gsol1}
\end{eqnarray}
where
\begin{eqnarray}
&&A=\sqrt{\epsilon_0^2(T_m=\infty) - g^2_{eff,0}(T_m=\infty)}\\
&&\epsilon_0(T_m=\infty) = K_{neq}-2 \label{A}
\end{eqnarray}

We solve the above equations upto a scale $l_1$=$T_{m0}\Lambda_0$.
Following this, for larger values of
$l$, the RG equations change due to a change in the scaling dimension and also because $I_K \sim T_m^2$.
In order to smoothly connect the solutions
of the RG equations at short and at long times we will make the ansatz,
$I_K\left(\frac{T_{m0}\Lambda_0}{l}\right) = I_K(\infty) \frac{\frac{T_{m0}^2 \Lambda_0^2}{l^2}}{1 + \frac{T_{m0}^2\Lambda_0^2}{l^2}}$
so that in the second step, the RG equations change
to  $\frac{d g_{eff}}{d\ln{l}}$=$-\left(\frac{\gamma^2K_0}{4}-2\right) g_{eff},
\frac{d\epsilon}{d\ln{l}}$=$- g_{eff}^2\frac{T_{m0}^2\Lambda_0^2}{l^2}$. The solution of these
equations with the initial conditions corresponding to Eqns.~\ref{esol1},~\ref{gsol1} evaluated at $l_1=T_{m0}\Lambda_0$  give,
\begin{eqnarray}
&&\epsilon^*(\Lambda_0T_{m0}\gg 1) = \frac{A}{\tanh\left[A\ln(\Lambda_0T_{m0})
+ \tanh^{-1}\left(\frac{A}{\epsilon_0}\right)\right]}\nonumber \\
&&\!\!\!
-\left(\frac{1}{\frac{\gamma^2K_0}{2}-2}\right)
\!\!\!
\left(\frac{A}{\sinh\left[A\ln(\Lambda_0T_{m0})
+ \tanh^{-1}\left(\frac{A}{\epsilon_0}\right)\right]}\right)^2\nonumber \\
\end{eqnarray}

Thus at long times a gapless phase is
recovered, where the interaction parameter $\epsilon^*$ reaches its steady-state value
of $A$ as a power-law with a non-universal exponent,
$\epsilon^* \xrightarrow{\Lambda_0T_{m0}\gg 1} A + {\cal O}\left(\frac{1}{\Lambda_0 T_{m0}}\right)^{2 A}$.

\subsection{Case (b)}
This occurs for $\epsilon(T_m)<0$. Thus
we are always in the strong coupling regime. Here we integrate the RG equations upto a scale
$l^*(T_m)$ where the renormalized coupling is {\cal O}(1). Beyond this
scale our RG equations are not valid, however the advantage of the bosonic theory is that at strong-coupling
$g\cos(\gamma \phi)\simeq g(1- \gamma^2\phi^2/2 + \ldots)$ so that $\sqrt{g}$ may be identified with a gap
or order-parameter.
The physical gap/order-parameter is then given by $\sqrt{g}/l^* = 1/l^*$.
Since $l^*(T_m)$ depends on time, it tells us how the order-parameter evolves in time.


At long times after the quench ($T_{m0}\Lambda_0 \gg 1$), the scaling dimension is determined by $K_{neq}$, while 
$g_{eff,0}$ may be approximated by its steady state value $g_{eff,0}(\infty)$.  
Here the RG equations are,
\begin{eqnarray}
\frac{d\epsilon}{d\ln{l}}\simeq 0\\
\frac{dg_{eff}}{d\ln{l}}= (-K_{neq}+2)g_{eff}
\end{eqnarray}
Integrating upto a scale where $g_{eff} \sim 1$, and provided that $\Delta T_{m0}\gg 1$, we find the steady-state
gap/order-parameter,
$\Delta_{ss}= \left(g_{eff,0}\right)^{\frac{1}{2-K_{neq}}}$.
Compare this with the gap in the
ground state of $H_f$~\cite{Giamarchibook} $\Delta_{eq}$=$\left(g_{eff,0}\right)^{\frac{1}{2-K_{eq}}}$.
Since $K_{neq}>K_{eq}$ and $g_{eff,0}\ll 1$, the gap at long times after the quench is always
smaller than the gap in equilibrium.

We now turn to intermediate times $\frac{1}{\Lambda_0}\ll T_{m0}\ll \frac{1}{\Delta}$, here
the RG involves two steps, the first in which the scaling dimension is determined
by $K_{neq}$ and the integration stops at $l_1=\Lambda_0T_{m0}$, and the second in which the scaling dimension
is determined by $K_0$, and the RG is terminated at $g_{eff}(l^*)=1$. Here we find (again taking $\frac{d\epsilon}{d\ln{l}}\simeq 0$),
$\Delta =
\left[
\left(\Lambda_0T_{m0}\right)^{\frac{\gamma^2K_0}{4}-K_{neq}}g_{eff,0}\right]^{\frac{1}{2-\frac{\gamma^2K_0}{4}}}
$.
Thus at intermediate times the order-parameter decreases with time if $K_{neq}> \frac{\gamma^2 K_0}{4}$,
or increases with time for the reverse case. Note that for $K_0$=$K$
this intermediate time power-law dynamics is absent.

\subsection{Case (c)}
{Case (c), periodic potential relevant at short times, and irrelevant at long times}.
This occurs for $\frac{\gamma^2K_0}{4}<2,K_{neq}>2$ and $g_{eff}$ not too large.
Here the short time behavior is the same as Case (b). At long times, the RG
proceeds in two steps. One where $T_{m0}\Lambda_0 > l > 1$ where the solution
is in equations~\ref{esol1},~\ref{gsol1}. For the second step the RG equations become
$\frac{dg_{eff}}{d\ln{l}} = g_{eff}(2-\frac{\gamma^2 K_0}{4}), \frac{d\epsilon}{d\ln{l}}\simeq 0$.
At the second step the RG equations are integrated upto a $l^*$ such that $g_{eff}(l^*)\sim 1$. The solution of the
order-parameter $\Delta \sim 1/l^*$ is found to be,
\begin{eqnarray}
&&\Delta(\Lambda_0T_{m0}\gg 1)=\nonumber \\
&&\left(\frac{1}{\Lambda_0 T_{m0}}\right)\left[\frac{A}{\sinh
\left[A\ln(\Lambda_0T_{m0})
+ \tanh^{-1}\left(\frac{A}{\epsilon_0}\right)\right]}\right]^\frac{1}{2-\frac{\gamma^2 K_0}{4}}\nonumber \\
\end{eqnarray}
Thus at long times, the order-parameter decreases to zero as $\Delta \sim
 \left(\frac{1}{\Lambda_0 T_{m0}}\right)^{1+\frac{A}{2-\frac{\gamma^2K_0}{4}}}$. This is of course only the contribution
 from the relevant perturbation. Due to irrelevant terms, the steady-state will still be characterized by
 a non-zero order-parameter.

\subsection{Case(d)}
Here we will consider the case of the dynamical transition where at short times the periodic potential is
irrelevant, whereas it becomes relevant at a finite non-zero time.
Let us for simplicity consider this scenario for a pure lattice quench so that $K_0=K$,
with $\frac{\gamma^2 K}{4} > 2$ such that $\epsilon >0$. This would naively imply that the periodic potential is irrelevant,
however that is not the case when $g_{eff}$ becomes large enough at some time $T_m^*$. We demonstrate below
how the dynamical transition occurs.

\begin{figure}
\centering
\includegraphics[totalheight=5cm]{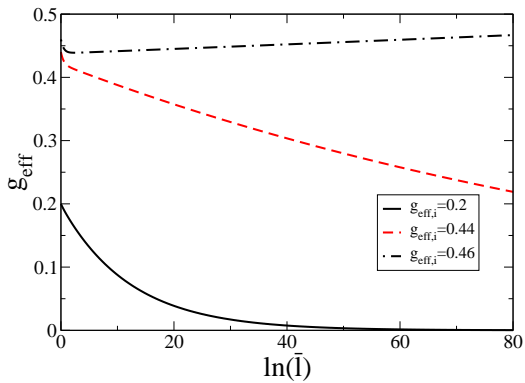}
\caption{RG flow of $g_{eff}(\bar{l}= \frac{l}{\Lambda_0 T_{m0}})$ for three different
initial conditions $g_{eff,i}=g_{eff}(\bar{l}=1)$
and hence times, and for an initial $\epsilon_i = \epsilon(\bar{l}=1)=0.1$. Critical coupling or time is located at
$g_{eff,c} = \sqrt{2\epsilon_i + \epsilon_i^2} = 0.458$.}
\label{gflow}
\end{figure}

Let us suppose we are at a time $T_{m0}\Lambda_0 \gg 1$.
At these times, the RG equations have to be solved in two steps. The first is when $1<l< T_{m0}\Lambda_0$. Here,
we may neglect the $l$-dependence of $I_K$ by noting that it changes slowly in time at long times, in particular
as a power-law, eventually reaching a steady-state.
So assuming $\frac{1}{2}|\frac{d\ln{I_{K}(l)}}{d\ln{l}}| \ll |\frac{d\ln{g}}{d\ln{l}}|$
in this regime, we get the following
conventional RG equations,
\begin{eqnarray}
\frac{d{g}_{eff}}{d\ln{l}} = -{g}_{eff}\epsilon \\
\frac{d\epsilon}{d\ln{l}} = -{g}_{eff}^2\\
\end{eqnarray}
where
\begin{eqnarray}
{g}_{eff,0} = g_0 \sqrt{I_K(\infty)\frac{\pi \gamma^2}{8}}\left(\frac{\gamma K}{2}\right)
\end{eqnarray}
The above equations may be solved by defining a constant of the flow
\begin{eqnarray}
D^2 \!\!= g_{eff}^2(l) -\epsilon^2(l) = g_{eff,0}^2 -\epsilon_0^2
\end{eqnarray}
Integrating upto $l = \Lambda_0 T_{m0}$ we get,
\begin{eqnarray}
&&\!\!\epsilon(l=\Lambda_0T_{m0})\!\! = \!\!D \tan\left[\arctan\left(\frac{\epsilon_0}{D}\right)-D \ln\left(\Lambda_0 T_{m0}\right)\right]\label{e11}\\
&&g_{eff}(l=\Lambda_0T_{m0}) = \sqrt{D^2 + \epsilon^2(l=\Lambda_0T_{m0})}\label{g11}
\end{eqnarray}
At the next step, the RG equations change because now $I_K \sim 1/l^2$. In order to smoothly connect the solutions
of the RG equations at short and at long times we will make the ansatz,
\begin{eqnarray}
I_K\left(\frac{T_{m0}\Lambda_0}{l}\right) = I_K(\infty)
\left[\frac{\frac{T_{m0}^2 \Lambda_0^2}{l^2}}{1 + \frac{T_{m0}^2\Lambda_0^2}{l^2}}\right]
\end{eqnarray}
The RG equations at the second step become,
\begin{eqnarray}
\frac{dg_{eff}}{d\ln{\bar{l}}} = -g_{eff}\epsilon \label{rgt1}\\
\frac{d\epsilon}{d\ln{\bar{l}}} = -\frac{g_{eff}^2}{\bar{l}^2} \label{rgt2}\\
\bar{l} = \frac{l}{\Lambda_0 T_{m0}}
\end{eqnarray}
with the initial conditions at $\bar{l}=1$ given by
Eqns~\ref{e11},~\ref{g11}.
Equations ~\ref{rgt1},~\ref{rgt2} can be solved exactly. The solution is
\begin{eqnarray}
\epsilon(\bar{l}) = -1 + \sqrt{a} \tan\left[\sqrt{a}\left(b-\ln{\bar{l}}\right)\right] \label{sol1}\\
g_{eff}(\bar{l}) = -\sqrt{a}\sec\left[\sqrt{a}\left(b-\ln{\bar{l}}\right)\right]e^{\ln{\bar{l}}} 
\label{sol2}
\end{eqnarray}
where
\begin{eqnarray}
a = g_{eff}^2(\bar{l}=1) - \left(\epsilon(\bar{l}=1)+1\right)^2\\
b = \frac{1}{\sqrt{a}}\tan^{-1}\left(\frac{\epsilon(\bar{l}=1) + 1}{\sqrt{a}}\right)
\end{eqnarray}
The solution for $g_{eff}(\bar{l})$ for different initial conditions are plotted in Fig.~\ref{gflow}. There is
a dynamical transition when $g_{eff}(\bar{l}=1) = g_{eff,c}$ where
\begin{eqnarray}
g_{eff,c} = \sqrt{\epsilon^2(\bar{l}=1) + 2\epsilon(\bar{l}=1)} \label{cond1}
\end{eqnarray}
The above critical coupling implies  that there is a dynamical transition
at a critical time $T_m^*$ such that
\begin{eqnarray}
\!\!g_{eff}(l=\Lambda_0T_{m}^*)\!\! = \sqrt{\epsilon^2(l=\Lambda_0T_{m}^*) + 2\epsilon(l=\Lambda_0 T_{m}^*)}
\end{eqnarray}
From Eq.~\ref{g11}, the above implies
\begin{eqnarray}
D^2 = 2\epsilon(l=\Lambda_0 T_m^*)
\end{eqnarray}
Using Eq.~\ref{e11}, we solve for $T_m^*$,
\begin{eqnarray}
\Lambda_0 T_m^* = e^{\frac{1}{D}\arctan\left(\frac{\epsilon_0}{D}\right)-\frac{1}{D}\arctan{\frac{D}{2}}}
\end{eqnarray}
As before, we may identify the order-parameter/gap as the length scale $l^*$ at which $g_{eff}({l}^*) \sim 1$. This
gives us the following result for how the order-parameter grows after the critical time,
\begin{eqnarray}
&&\Delta = \frac{1}{l^*}\nonumber \\
&&=\theta(T_{m0}-T_m^*)\frac{1}{\Lambda_0 T_{m0}}\left[g_{eff}(l=\Lambda_0T_{m0})\right]^
{\frac{1}{\epsilon(l=T_m^*)-\epsilon(l=T_{m0})}}\nonumber  \\
&&\sim \theta(T_{m0}-T_m^*)\frac{1}{\Lambda_0 T_{m0}}\left[g_{eff}(l=\Lambda_0T_{m0})\right]^{\frac{f_2}{(T_{m0}-T_m^*)}}
\end{eqnarray}
where $f_2=\frac{1}{|\frac{d\epsilon(l=T_m)}{dT_m}|_{T_m=T_m^*}|}$. 
This is the contribution from the relevant perturbation. However irrelevant terms of the form $\cos(2\gamma\phi), \cos(3\gamma\phi)$
etc are always present that give a non-zero contribution to the order-parameter which we denote as $\Delta_{smooth}$. 
These may be evaluated perturbatively, and 
correspond to a smooth behavior in time. Due to these terms, the order-parameter is never strictly speaking zero, however
as shown above,  the
relevant term causes the time-evolution
of the order-parameter to be non-analytic in time as it grows as $e^{-c/(T_m-T_m^*)}$ after a critical time $T_m^*$.
We identify this non-analyticity with a dynamical phase transition.

\begin{thebibliography}{29}
\bibitem [{not()}]{noteLE}%
  \BibitemOpen
  \href@noop {} {}\bibinfo {note} {For a pure lattice quench at the
  Luther-Emery point ($K$=$K_0$,$K_{eq}$=$1$), at short times, $I_K \sim{\cal
  O}(T_m^4)$, so that $g_{eff}\sim T_m^2$.}\BibitemShut {Stop}%
\bibitem [{not()}]{suppRG2}%
  \BibitemOpen
  \href@noop {} {}\bibinfo {note} {$\Delta_m$ is related to $\Delta(T_m) $=$\frac{1}{l^*(T_m)}$ as 
$\Delta_m(T_m) \sim \left[\Delta(T_m)\right]^{m^2 K_{neq}} \forall\,\, T_m \gg 1$.
}\BibitemShut {Stop}%
\end{thebibliography}

%

\end{document}